\documentclass[twocolumn,eqsecnum,showpacs,pra]{revtex4}
\usepackage{graphicx,amssymb,bm}
\usepackage{color}

\newcommand{\beq}{\begin{equation}}
\newcommand{\eeq}{\end{equation}}
\newcommand{\beqa}{\begin{eqnarray}}
\newcommand{\eeqa}{\end{eqnarray}}

\def\ket#1{|\,#1\,\rangle}
\def\bra#1{\langle\, #1\,|}

\def\proj#1#2{\ket{#1}\bra{#2}}

\def\opone{\leavevmode\hbox{\small1\kern-3.8pt\normalsize1}}

\newcommand{\pc}{\mathbb{P}_{\rm C}}

\begin{document}
\title{Probing light polarization with the quantum Chernoff bound}

\author{Iulia Ghiu$^{1}$}

\author{Gunnar Bj\"{o}rk$^2$}

\author{Paulina Marian$^{1,3}$}

\author{Tudor A. Marian$^1$}

\affiliation{$^1$ Centre for Advanced Quantum Physics,\\
Department of Physics, University of Bucharest,\\
P. O.  Box MG-11, R-077125  Bucharest-M\u{a}gurele, Romania}
\affiliation{$^2$ School of Information and Communication Technology, Royal Institute of Technology (KTH),
Electrum 229, SE-164 40 Kista, Sweden}
\affiliation{$^3$  Department of Chemistry, University of Bucharest,\\
Boulevard Regina Elisabeta 4-12, R-030018  Bucharest, Romania}

\date{\today}

\begin{abstract}
We recall the framework of a consistent quantum description of polarization
of light. Accordingly, the degree of polarization of a two-mode state
$\hat \rho$ of the quantum radiation field can be defined as a distance
of a related state ${\hat \rho}_b$ to the convex set of all SU(2) invariant
two-mode states. We explore a distance-type polarization measure in terms
of the quantum Chernoff bound and derive its explicit expression. A comparison between the Chernoff and Bures degrees of polarization leads to interesting conclusions for some particular states chosen
as illustrative examples.
\end{abstract}

\maketitle

\section{Introduction}

Polarized states of the quantum electromagnetic field are basic resources
in many experiments in quantum optics and quantum information processing,
e.g., Bell inequalities \cite{Kwiat}, quantum tomography \cite{Barbieri},
quantum cryptography \cite{Bennett1,Muller}, quantum teleportation
\cite{Bennett2, Bouw}, superdense coding \cite{Bennett3},
entanglement swapping \cite{Bose}, entanglement purification for quantum communication \cite{Pan}, and quantum computation \cite{Joo}.

In classical optics, the degree of polarization is defined in terms
of the Stokes parameters \cite{Mandel}. The classical definition
was adapted to quantum optics, where the Stokes parameters have been
replaced by the expectation values of the Stokes operators \cite{Simon}.
However, this polarization measure contains only second-order
correlations of the field, which are not sufficient for a complete
description of all quantum-optics problems, where higher-order correlations
play an important role. An idea to eliminate this drawback is due to Luis,
who quantified the polarization in terms of the variance over $S^2$
of the SU(2) $Q$ function for the given field state \cite{Luis1,Luis2,Luis3}.
Alternatively, the degree of polarization has been defined as the minimal
overlap between the given state and any state obtained from it via
a SU(2) transformation \cite{Bjork1,Sehat}. Other attempts have been made
to introduce a polarization measure for electromagnetic near fields
by using the Gell-Mann matrices \cite{Setala1,Setala2,Vitanov}.
Recently, the degree of polarization has been defined as a distance between
the field state in question and the set of unpolarized states. Several metrics,
e.g., the Hilbert-Schmidt and Bures metrics, have been used for evaluating
the polarization of some field states  \cite{Bjork2,Bjork3,Bjork4}.

In this work we introduce a distance-type degree of polarization defined
in terms of the quantum Chernoff bound. In a seminal paper, Chernoff
investigated the problem of discriminating two probability distributions
and found an upper bound on the minimal error probability $P_{\rm min}^{(N)}$
in the asymptotic case $(N\to \infty)$ \cite{Chernoff}. This is known as the classical Chernoff bound and has many applications in statistical decision theory.
After some 55 years, this bound was generalized to the quantum case.
First, Ogawa and Hayashi proposed three promising candidates for a quantum expression \cite{Ogawa}. After some other subsequent progress \cite{Kargin}, the quantum Chernoff bound was proven to coincide with one
of their formulas. This important result was established through the conjugate efforts of two groups of researchers: Nussbaum and Szko\l a,
and Audenaert {\it et al.} \cite{Nussbaum,Acin}.
The quantum scenario is as follows: $N$ identical copies of a quantum system
are prepared in the same unknown state, which is either $\hat \rho$ or
$\hat \sigma$. The task at hand is to determine the minimal probability of error
by testing the copies in order to draw a conclusion about the identity
of the state. When the two states are equiprobable, the minimal error
probability of discriminating them in a measurement performed
on $N$ independent copies is \cite{Kargin,Auden}
\begin{equation}
P_{\rm min}^{(N)}(\hat \rho,\, \hat \sigma)=\frac{1}{2}\left(1-\frac{1}{2}
||{\hat \rho}^{\otimes N}-{\hat \sigma}^{\otimes N}||_1\right)  \label{error},
\end{equation}
where $||\hat A||_1:=\mbox{Tr}\sqrt{{\hat A}^{\dagger}\hat A}$ is
the trace norm of a trace-class operator $\hat A.$  In the special case
when both states are pure (denoted by $|\Phi \rangle$ and $|\Psi \rangle$),
the minimal error probability\ (\ref{error}) reads \cite{Kargin}
$$P_{\rm min}^{(N)}(|\Phi \rangle \langle\Phi|, \;|\Psi \rangle \langle\Psi|)
=\frac{1}{2}\left( 1-\sqrt{1-|\langle \Phi \ket{\Psi}|^{2N}}\right).$$
For an optimal asymptotic testing  $(N\to \infty )$, an upper bound
$P^{(N)}_{QCB}$ of the minimal probability of error \ (\ref{error})
was found to decrease exponentially with $N$ \cite{Acin,Auden}:
$$P^{(N)}_{QCB}(\hat \rho,\, \hat \sigma)\sim\exp \left[-N
\xi _{QCB}(\hat \rho, \,\hat \sigma)\right], \quad\qquad (N\gg 1),$$
where the positive quantity
\begin{equation}
\xi_{QCB}(\hat \rho,\, \hat \sigma):=-\ln \left[ \min_{s\in [0,1]}
\mbox{Tr}\left({\hat \rho}^s{\hat \sigma}^{1-s}\right) \right]  \label{xi}
\end{equation}
is called {\em quantum Chernoff bound} \cite{Nussbaum,Acin,Auden}.

We find it convenient to introduce the function
\begin{equation} Q(\hat \rho,\, \hat \sigma):=\min_{s\in [0,1]}
\mbox{Tr}({\hat \rho}^s{\hat \sigma}^{1-s}), \label{QCB}
\end{equation}
which is manifestly symmetric, $Q(\hat \rho,\, \hat \sigma)
=Q(\hat \sigma,\,\hat \rho),$ and is referred to in what follows
as the quantum Chernoff overlap of the states $\hat \rho$
and $\hat \sigma$ \cite{Marian}.
Its maximal value is reached when the states $\hat \rho$
and $\hat \sigma$ coincide.
In the body of the paper we intensively employ the quantities
\begin{equation}
Q_s(\hat \rho,\, \hat \sigma):={\rm Tr}({\hat \rho}^s{\hat \sigma}^{1-s}),
\label{Renyi}
\end{equation}
which are the quantum analogues of the classical R\'enyi overlaps
discussed in Ref. \cite{Fuchs} as being distinguishability
measures in their own right. According to Eqs.\ (\ref{xi}) and\ (\ref{QCB}),
their minimum over $s$ determines the quantum Chernoff bound, which
has many applications in various branches of physics. Calsamiglia
{\it et al.} have employed it as a  measure of distinguishability
between qubit states and between single-mode Gaussian states of the
radiation field \cite{Calsa}. Hiai {\it et al.} have analyzed
the asymptotic discrimination of two states with measurements that are
invariant under some symmetry group of the system \cite{Hiai}. Recently,
the quantum Chernoff overlap was employed to evaluate the degree
of non-classicality for one-mode Gaussian states \cite{Marian}.
Pirandola and Lloyd have found upper bounds for the error probability of discrimination of Gaussian states of $n$ bosonic modes \cite{Pirandola}.
They combined Minkowski's  inequality and the quantum Chernoff bound
and derived computable bounds. The quantum Chernoff bound was used
for asymptotic discrimination between two states of an infinite-lattice
system in the fermionic case \cite{Mosonyi1},
as well as in the bosonic one \cite{Mosonyi2}. The quantum Chernoff bound
is also applied to the theory of quantum phase transitions.
Abasto {\it et al.} have evaluated the quantum Chernoff metric
for the $XY$ model at finite temperature \cite{Abasto}.
By use of the quantum Chernoff bound, discrimination between two ground
states or two thermal states of the one-dimensional quantum Ising model
was recently addressed by Invernizzi and Paris \cite{Paris}.

The present article deals with two-mode states of the quantum radiation field.
Its purpose is to investigate a distance-type degree of polarization
that involves the quantum Chernoff overlap. The paper is organized as follows.
In Sec. II we review the recently formulated requirements to be
fulfilled by any acceptable measure of polarization \cite{Luis3,Bjork2}.
We here insist on the physical significance of these general
requirements that change the current view on the way
of evaluating the degree of polarization for a two-mode state.
Section III is devoted to the Chernoff degree of polarization
for which a general formula is derived and discussed. A parallel
treatment of the Bures degree of polarization is then presented.
In Sec. IV the obtained formulas are specialized to pure states.
The Chernoff and Bures degrees of polarization are compared
for two families of states, each of them having just two nonvanishing
photon-number probabilities. Our conclusions are outlined in Sec. V.

\section{Quantum degree of polarization}

The polarization transformations are an essential ingredient in linear
optics. They are carried out by lossless linear optical devices while
transmitting a quasimonochromatic light beam between a pair of planes
transverse to its travel direction. We give here two examples.
The first one is that of a compensator which introduces a phase
difference between two perpendicular components of the oscillating
electric field. A second device to be mentioned is called rotator
because it produces a rotation of the electric field vector about
the beam propagation axis.

From the mathematical point of view, the class of linear polarization
transformations is a group of unitary operators ${\hat U}_{\rm pol}$
on the two-mode Hilbert space ${\cal H}_{H}\otimes {\cal H}_{V}$.
They are generated by three Stokes operators:
\beqa
\hat S_1:&=&\hat a _{H}^\dagger \hat a_{V}+\hat a _{H} \hat a_{V}^\dagger ,
\qquad \hat S_2:=\frac{1}{i} \left( \hat a _{H}^\dagger \hat a_{V}
-\hat a _{H} \hat a_{V}^\dagger \right),\nonumber\\
\hat S_3:&=&
\hat a _{H}^\dagger \hat a_{H}-\hat a _{V}^\dagger \hat a_{V},
\label{Stokes}
\eeqa
built with the amplitude operators of the horizontal $(H)$ and vertical $(V)$
modes. Accordingly, the operators ${\hat U}_{\rm pol}$ form an infinite-dimensional unitary representation of the group SU(2) and can be parametrized in terms
of the Euler angles $\phi,\, \theta,\, \psi,\,$ as follows:
\beqa
{\hat U}_{\rm pol}(\phi,\,\theta,\,\psi)&=&\exp{\left(-i\, \frac{\phi}{2}\, \hat S_3\right)}
\exp{\left(-i\, \frac{\theta}{2}\, \hat S_2\right)}\nonumber\\
&&\times \exp{\left(-i\, \frac{\psi}{2}\, \hat S_3\right)}.
\label{Upol}
\eeqa
Any SU(2) polarization transformation (\ref{Upol}) preserves
the total number of photons, which is described by the fourth Stokes operator,
\begin{equation}
\hat S_0:=\hat a_{H}^\dagger \hat a_{H}+\hat a_{V}^\dagger \hat a_{V}.
\label{S0}
\end{equation}

A state $\hat \tau$ that remains invariant under any polarization
transformation (\ref{Upol}) is unpolarized \cite{Prakash}. It is known
for a long time that a  two-mode state $\hat \tau$ is SU(2) invariant
if and only if it has the spectral decomposition
\cite{Prakash,Agarwal,Lehner,Bjork5}
\begin{equation}
\hat \tau =\sum_{N=0}^\infty \; \pi_N \frac{1}{N+1}\, {\hat P}_N, \label{tau}
\end{equation}
where
\begin{equation}
{\hat P}_N:=\sum_{n=0}^N |n,N-n \rangle \langle n,N-n| \label{projection}
\end{equation}
is the projection operator onto the vector subspace of the $N$-photon states,
called the $N$th excitation manifold. We have denoted
$|n,N-n \rangle:={|n \rangle}_H\otimes {|N-n \rangle}_V$.
Further, $\pi_N$ are the photon-number probabilities in the SU(2) invariant
state $\hat \tau$ and they satisfy the normalization condition
\begin{equation}
\sum_{N=0}^{\infty}\pi_N=1. \label{certainty}
\end{equation}

In order to describe the polarization properties of an arbitrary two-mode
state $\hat \rho$, we make use of its photon-number-ordered Fock expansion
\beqa
\hat \rho &=&\sum_{M=0}^{\infty}\sum_{N=0}^{\infty}\sum_{m=0}^M\sum_{n=0}^N
|m,M-m \rangle \nonumber\\
&&\times \langle m,M-m|\hat \rho|n,N-n \rangle \langle n,N-n|.
\label{Fock}
\eeqa
The above expansion can be split into the sum of the block-diagonal terms
$(M=N)$ and that of the off-block-diagonal ones $(M\not=N)$. The former sum
is the block-diagonal density matrix ${\hat \rho}_b$
associated with the given state $\hat \rho$,
\begin{equation}
{\hat \rho}_b:=\sum_{N=0}^\infty p_N{\hat \rho}_N.
\label{rhob}
\end{equation}
In Eq.\ (\ref{rhob}), $p_N$ is the probability of the $N$th excitation manifold:
\begin{equation}
p_N={\rm Tr} ({\hat \rho}{\hat P}_N)=\sum_{n=0}^N\rho_{nn}^{(N)},
\label{N-probab}
\end{equation}
where
\begin{equation}
\rho_{mn}^{(N)}:=\langle m,N-m|\hat \rho|n,N-n \rangle, \label{N-dm}
\end{equation}
are the entries of a positive semidefinite matrix ${\rho}^{(N)}
\in {\mathcal M}_{N+1}(\mathbb{C}).$ Further, ${\hat \rho}_N$
is a $N$-photon state determined by the matrix ${\rho}^{(N)}$ with a nonvanishing trace $p_N$:

\beqa
{\hat \rho}_N:&=& \frac{1}{p_N}{\hat P}_N{\hat \rho}{\hat P}_N
=\frac{1}{p_N}\sum_{m=0}^N\sum_{n=0}^N|m,N-m \rangle \rho_{mn}^{(N)}\nonumber \\
&&\times \langle n,N-n|, \qquad p_N>0. \label{rhoN}
\eeqa

Recall now the requirements we need to quantify the polarization
of a two-mode state $\hat \rho$. There are three conditions to be satisfied
by its degree of polarization $\mathbb{P}(\hat \rho)$  \cite{Bjork4}:
\begin{description}
\item a) $\mathbb{P}(\hat \rho)=0$ if and only if $\hat \rho$
is unpolarized. This is only natural: for an unpolarized state the degree
of polarization vanishes and, conversely, a state with zero degree
of polarization is unpolarized.

\item
b) The degree of polarization is invariant under polarization transformations:
\begin{equation}
\mathbb{P}({\hat U}_{\rm pol}\,{\hat \rho}\,{\hat U}_{\rm pol}^{\dagger})
=\mathbb{P}(\hat \rho) \label{invariance}.
\end{equation}

\item
c) The degree of polarization is not affected by coherences between
different excitation manifolds. In fact, all polarization properties
of a given two-mode state $\hat \rho$ are not influenced by its coherent terms
between vector subspaces with different numbers of photons, displayed
in Eq.\ (\ref{Fock}). Excluding them, we ascribe the description
of polarization to the block-diagonal density matrix ${\hat \rho}_b$,
Eq.\ (\ref{rhob}). Accordingly,  we adopt a new definition for the degree
of polarization of the state (\ref{Fock}):
\begin{equation}
\mathbb{P}(\hat \rho):=\mathbb{P}({\hat \rho}_b). \label{degree}
\end{equation}
\end{description}

Equation\ (\ref{degree}) implies that all two-mode states with the same
block-diagonal part ${\hat \rho}_b$ are equally polarized. In particular,
any unpolarized state $\hat \sigma$ has an SU(2)-invariant
block-diagonal part ${\hat \sigma}_b$ \cite{Prakash}:
\begin{equation}
{\hat \sigma}_b =\sum_{N=0}^\infty \; \pi_N \frac{1}{N+1}\, {\hat P}_N.
\label{sigmab}
\end{equation}
We refer here only to type I unpolarized light \cite{Lehner}.
Note that, except for the vacuum, any unpolarized state is mixed.

The block-diagonal state ${\hat \rho}_b$ occurring in definition (\ref{degree})
has a significant operational meaning. Indeed, the observable (\ref{S0}),
\begin{equation}
\hat N:=\hat N_{H}+\hat N_{V}=\sum_{N=0}^\infty N\hat P_N,
\end{equation}
is a random variable that commutes with any polarization transformation
${\hat U}_{\rm pol}$. Consequently, a polarization measurement
of an arbitrary state does not alter its photon-number distribution. Now,
when we perform a von Neumann measurement of the total number of photons,
we obtain the outcome $N$ with the probability $p_N$, while the state
$\hat \rho$ collapses into the $N$-photon state ${\hat \rho}_N$,
Eq.\ (\ref{rhoN}). We measure the total number of photons for each
member of an ensemble of identical states described by $\hat \rho$
and do not select any result. In this way, we eventually get another
ensemble of states described by the mixture
${\hat \rho}_b=\sum_{N=0}^\infty p_N{\hat \rho}_N$.
Note that the block-diagonal state ${\hat \rho}_b$ has the same photon-number
distribution as the given state $\hat \rho$. This happens because
${\hat \rho}_b$ is deliberately built with the ensemble of states provided
by the corresponding von Neumann measurement. To sum up, an ideal
non-selective measurement of the total number of photons is a quantum
operation \cite{Nielsen} (or quantum channel) ${\cal B}$  whose output
is ${\hat \rho}_b$:
\begin{equation}
\hat \rho \stackrel{{\cal B}}{\longrightarrow } {\hat \rho}_b=
\sum_{N=0}^\infty {\hat P}_N{\hat \rho}{\hat P}_N \label{channel}
\end{equation}
The quantum operation ${\cal B}$ preserves the photon-number distribution.
Remark first that any output ${\hat \rho}_b$ of the channel ${\cal B}$
commutes with the output ${\hat \sigma}_b$, Eq.\ (\ref{sigmab}),
of an arbitrary unpolarized state $\hat \sigma$:
\begin{equation}
[{\hat \rho}_b,{\hat \sigma}_b]=0. \label{com}
\end{equation}
This is not generally true for the input states $\hat \rho$ and $\hat \sigma$.
As a consequence of the commutation relation\ (\ref{com}), most
polarization-measure candidates  $\mathbb{P}({\hat \rho}_b)$ depend only
on the photon-number probabilities $p_N$  and the eigenvalues
$\lambda_{N,n}$ of the density matrices $\frac{1}{p_N}\, {\rho}^{(N)}$
that determine the $N$-photon states ${\hat \rho}_N$ entering the convex
decomposition \ (\ref{rhob}). Since all these quantities are SU(2) invariant,
it follows that the candidates themselves fulfill the SU(2)-invariance
condition\ (\ref{invariance}) and are therefore admissible as adequate
measures of polarization \cite{Bjork2}.

\section{Chernoff degree of polarization}

\subsection{Definition}

In view of its outstanding distinguishability properties,
the quantum Chernoff bound can be used to define a polarization
measure similar to other distance-type ones \cite{Bjork2,Bjork4}.
We therefore introduce the Chernoff degree of polarization
\begin{equation}
\mathbb{P}_{\rm C}(\hat \rho):=1-\max_{\hat \sigma \in {\cal U}}
Q({\hat \rho}_b, \,\hat {\sigma}_b), \label{p-chern}
\end{equation}
built with the Chernoff overlap (\ref{QCB}). Here ${\hat \rho}_b$
is the block-diagonal state\ (\ref{rhob}) and ${\cal U}$ stands for
the set of all unpolarized two-mode states. Let us denote
\begin{equation}
\tilde Q :=\max_{\hat \sigma \in {\cal U}}Q({\hat \rho}_b,\,{\hat \sigma}_b),
\end{equation}
in order to write simply: $\mathbb{P}_{\rm C}(\hat \rho)=1-\tilde Q.$

It is important to check that definition (\ref{p-chern}) fulfills the three requirements stated in Sec. II. The  "if" part of property a) is obvious,
so that we are left to prove its  "only if" part.

To this end, let us consider an arbitrary block-diagonal state ${\hat \rho}_b$  which is polarized. As already mentioned, we have denoted by ${\lambda}_{N,n}$ the eigenvalues of any $N$-photon density matrix $\frac{1}{p_N}\, \rho^{(N)}$, $(p_N>0)$. Let $\nu_N$ be the rank of the matrix $\rho^{(N)}$, Eq.\ (\ref{N-dm}),
i.e., the number of its positive eigenvalues $p_N {\lambda}_{N,n}$:
\beqa
&&\nu_N:={\rm rank}\,{\rho}^{(N)},  \qquad {\rho}^{(N)}
\in {\mathcal M}_{N+1}(\mathbb{C}), \nonumber \\
&&1 \leqq {\nu_N} \leqq {N+1}.
\label{rank}
\eeqa
For subsequent use, we introduce the quantity
\begin{equation}
{\xi}_N^{(s)}:=\sum_{n=0}^N(\lambda_{N,n})^s, \qquad p_N>0,
\label{xiN}
\end{equation}
which is a decreasing function of $s$ from the limit ${\xi}_N^{(0)}=\nu_N$
to the value ${\xi}_N^{(1)}=1.$

The commuting density operators  ${\hat \rho}_b,$ Eq.\ (\ref{rhob}), and
${\hat \sigma}_b,$ Eq.\ (\ref{sigmab}), have the eigenvalues $p_N \lambda_{N,n}$
and $\pi_N \frac{\delta_{nn}}{N+1}$, respectively. Therefore, a R\'enyi
overlap of the states ${\hat \rho}_b$  and ${\hat \sigma}_b$ reads
\beqa
Q_s({\hat \rho}_b, \,{\hat \sigma}_b)&=&\sum_{N=0}^\infty \sum_{n=0}^N \:
(p_N \lambda_{N,n})^s\left(\pi_N \frac{\delta_{nn}}{N+1} \right)^{1-s}, \nonumber \\
&&\qquad 0 \leqq s \leqq 1.  \label{Renyi1}
\eeqa
Obviously,
\begin{equation}
Q_0({\hat \rho}_b, \,{\hat \sigma}_b) \leqq 1, \qquad \qquad
Q_1({\hat \rho}_b, \,{\hat \sigma}_b) \leqq 1.  \label{ends}
\end{equation}
For $0<s<1$, we apply H\"older's inequality \cite{HLP}:
\begin{equation}
\sum_{n}a_n b_n\leqq \left[\sum_{m}(a_m)^{\,p}\right]^{\frac{1}{p}}
\left[\sum_{n}(b_n)^{\,q}\right]^{\frac{1}{q}}.
\label{Hoelder}
\end{equation}
In Eq. (\ref{Hoelder}), $a_n \geqq 0, \; b_n \geqq 0,$ and $\{p,\; q\}$
is a pair of {\em conjugate exponents}, i.e., positive real numbers
such that $\,p+q=pq\,$ or, equivalently, $\, \frac{1}{p}+\frac{1}{q}=1.\,$
Equation (\ref{Hoelder}) becomes an equality if and only if $a_n$ and $b_n$
are components of proportional vectors. When $p$ is conjugate to itself
$\,(p=q=2)\,$, H\"{o}lder's inequality\ (\ref{Hoelder}) reduces to Cauchy's inequality.

We specialize Eq. (\ref{Hoelder}) by taking
$$a_{N,n}=(p_N \lambda_{N,n})^s,
\quad b_{N,n}=\left(\pi_N \frac{\delta_{nn}}{N+1} \right)^{1-s},$$
$$\qquad p=\frac{1}{s}\,, \quad q=\frac{1}{1-s}\,, \qquad 0<s<1,$$
to get the inequality:
\beqa
&&Q_s({\hat \rho}_b,\,{\hat \sigma}_b)<\left(\sum_{M=0}^{\infty}\sum_{m=0}^M
p_M \lambda_{M,m}\right)^s\nonumber \\
&&\times \left(\sum_{N=0}^{\infty}\sum_{n=0}^N
\pi_N \frac{\delta_{nn}}{N+1} \right)^{1-s}=1,
 \hspace{0.15cm} 0<s<1.
\label{Qs<1}
\eeqa
In Eq.\ (\ref{Qs<1}), a strict inequality holds because the states
${\hat \rho}_b$ and ${\hat \sigma}_b$ cannot coincide: the first one
is polarized and the second is not. The same strict inequality
is still valid for the maximum of the R\'enyi overlap occurring
in Eq.\ (\ref{Qs<1}):
\begin{equation}
\max_{\hat \sigma \in \mathcal U} Q_s({\hat \rho}_b, \,{\hat \sigma}_b)<1,
\qquad 0<s<1. \label{maxQs<1}
\end{equation}
Taking into account the identity
\begin{equation}
\max_{\hat \sigma \in {\cal U}}\min_{s\in [0,1]}
Q_s({\hat \rho}_b, \,{\hat \sigma}_b)
=\min_{s\in [0,1]}\max_{\hat \sigma \in {\cal U}}
Q_s({\hat \rho}_b, \,{\hat \sigma}_b), \label{minmax}
\end{equation}
an inspection of Eqs.\ (\ref{ends}) and\ (\ref{maxQs<1}) leads to
the inequality to be proven:
\begin{equation}
\mathbb{P}_{\rm C}(\hat \rho)=1-\tilde Q>0. \label{polariz}
\end{equation}
Equation\ (\ref{polariz}) is then true for any state $\hat \rho$
whose block-diagonal part ${\hat \rho}_b$ is polarized.

Property b) is immediate. Indeed, any polarization transformation
${\hat U}_{\rm pol}$ is the orthogonal sum of all the SU(2) irreducible
representations and their carrier spaces are just the corresponding
$N$-photon eigensubspaces. Consequently, the block-diagonal part
of the state ${\hat U}_{\rm pol}\,{\hat \rho}\,{\hat U}_{\rm pol}^{\dagger}$
factors as follows:
\begin{equation}
\left({\hat U}_{\rm pol}\,{\hat \rho}\,{\hat U}_{\rm pol}^{\dagger} \right)_b
={\hat U}_{\rm pol}\,{\hat \rho}_b\,{\hat U}_{\rm pol}^{\dagger}.
\end{equation}
By use of the invariance of the Chernoff overlap under unitary
transformations \cite{Auden},
$$Q(\hat U {\hat \rho}_1 \hat U^\dagger,\, \hat U {\hat \rho}_2 \hat U^\dagger)
=Q({\hat \rho}_1,\, {\hat \rho}_2),$$
we get
\begin{equation}
Q({\hat U}_{\rm pol}\,{\hat \rho}_b\,{\hat U}_{\rm pol}^{\dagger}, \,
{\hat \sigma}_b)=Q({\hat \rho}_b, \,{\hat U}_{\rm pol}^{\dagger} \,
{\hat \sigma}_b\,{\hat U}_{\rm pol})=Q({\hat \rho}_b, \,{\hat \sigma}_b).
\label{invar}
\end{equation}
The last equality in Eq.\ (\ref{invar}) follows from the SU(2) invariant
formula\ (\ref{sigmab}) corresponding to any unpolarized two-mode state
$\hat \sigma$. Hence we obtain the SU(2) invariance property
\begin{equation}
\mathbb{P}_{\rm C}({\hat U}_{\rm pol}\,{\hat \rho}\,{\hat U}_{\rm pol}^{\dagger})
=\mathbb{P}_{\rm C}(\hat \rho). \label{p2}
\end{equation}

Property c) is fulfilled by definition.

\subsection{General expression}

Our task here is to evaluate the parameters $\tilde \pi_N$ of the unpolarized
state for which the maximum in Eq.\ (\ref{p-chern}) is obtained.
Determining  $\tilde Q$ is equivalent to finding the saddle point
of the \ function $Q_s({\hat \rho}_b, \,{\hat \sigma}_b)$. We start
by writing the R\'enyi overlap $Q_s({\hat \rho}_b, \,{\hat \sigma}_b)$,
Eq.\ (\ref{Renyi1}), in an equivalent form:
\begin{equation}
Q_s({\hat \rho}_b, \,{\hat \sigma}_b)=\sum_{N=0}^\infty \:(p_N)^s \xi_N^{(s)}
\left(\frac{\pi_N }{N+1} \right)^{1-s},  \hspace{0.5cm} 0 \leqq s \leqq 1.
\label{Renyi2}
\end{equation}
Let us treat first the case $s>0$. The maximum of the R\'enyi overlap
$Q_s({\hat \rho}_b, \,{\hat \sigma}_b)$ with respect to the variables $\pi_N$
under the constraint (\ref{certainty}) can be found by applying the method
of the Lagrange multipliers. One readily gets the $N$-photon probabilities
${\tilde \pi_N}^{(s)}$ that maximize the function (\ref{Renyi2}):
\begin{equation}
\tilde \pi_N^{(s)}=\frac{p_N\left( \xi_N^{(s)}\right) ^{1/s}(N+1)^{1-\frac{1}{s}}}
{\sum_{M=0}^\infty p_M\left( \xi_M^{(s)}\right) ^{1/s}(M+1)^{1-\frac{1}{s}}}.
\label{par}
\end{equation}
They characterize the closest unpolarized state ${\hat {\tilde \sigma}}_b$
to the state $\hat {\rho}_b$,
\begin{equation}
Q_s({\hat \rho}_b, \,\hat {\tilde \sigma}_b):=\max_{{\hat \sigma}
\in {\cal U}}\,Q_s({\hat \rho}_b, \,{\hat \sigma}_b).
\label{close}
\end{equation}
Insertion of Eq.\ (\ref{par}) into Eq.\ (\ref{Renyi2}) gives the explicit
formula
\beqa
Q_s({\hat \rho}_b, \,\hat {\tilde \sigma}_b)&=&\left\{ \sum_{N=0}^\infty p_N(N+1)
\left[\frac{\xi_N^{(s)}}{N+1}\right]^{1/s} \right\}^s, \nonumber \\
&&0<s\leqq 1. \label{explicit}
\eeqa
It is convenient to denote by $\tilde {N}(s)$ the value of $N$
that maximizes the ratio $\frac{{\xi}_N^{(s)}}{N+1}:$
\begin{equation}
\frac{{\xi}_{\tilde {N}(s)}^{(s)}}{\tilde {N}(s)+1}:=\max_{0\leqq N<\infty}
\frac{{\xi}_N^{(s)}}{N+1}.
\label{tildeNs}
\end{equation}
Equations (\ref{explicit}) and (\ref{tildeNs}) imply the inequality
\begin{equation}
\max_{\hat \sigma \in \mathcal U} Q_s({\hat \rho}_b, \,{\hat \sigma}_b)
\leqq \frac{{\xi}_{\tilde {N}(s)}^{(s)}}{\tilde {N}(s)+1}\,
(\langle N\rangle+1)^s, \;\; \qquad  0<s\leqq 1. \label{in-s-gen}
\end{equation}

We are now ready to handle the limit case $s=0$. Recalling that
${\xi}_N^{(0)}=\nu_N$ and setting $\tilde {N}:=\tilde {N}(0),\,$
Eq.\ (\ref{tildeNs}) reads for $s=0$
\begin{equation}
\frac{{\nu}_{\tilde N}}{\tilde {N}+1}:=\max_{0\leqq N<\infty}
\frac{{\nu}_N}{N+1}. \label{tildeN}
\end{equation}
The inequality (\ref{in-s-gen}) has therefore the limit
\begin{equation}
\lim_{s\to 0}Q_s({\hat \rho}_b, \,{\hat {\tilde \sigma}}_b)
\leqq\frac{{\nu}_{\tilde N}}{\tilde {N}+1} \label{lim}.
\end{equation}
If we consider the unpolarized $\tilde N$-photon state
\begin{equation}
{\hat \sigma}_{\tilde N}=\frac{1}{\tilde {N}+1}{\hat P}_{\tilde N},
\label{unpolarized}
\end{equation}
i.e., with $\pi_N=\delta_{N, \tilde N}$, then, according to
Eq.\ (\ref{Renyi2}) we get
\begin{equation}
Q_s(\hat \rho_b,\,{\hat \sigma}_{\tilde N})
=(p_{\tilde N})^s{\xi}_{\tilde N}^{(s)}\left(\frac{1}{\tilde {N}+1}
\right)^{1-s}, \;\; \qquad 0\leqq s\leqq 1. \label{N-photon_s}
\end{equation}
The limit $s=0$ of Eq. (\ref{N-photon_s}) then reads
\begin{equation}
\lim_{s\to 0}Q_s({\hat \rho}_b, \,{\hat \sigma}_{\tilde N})
=\frac{{\nu}_{\tilde N}}{\tilde {N}+1}.
\label{N-photon_0}
\end{equation}
Equations (\ref{lim}) and  (\ref{N-photon_0}) show that for $s=0$ the
unpolarized state (\ref{unpolarized}) is the closest to $\hat \rho_b$.
Therefore, the explicit formula (\ref{explicit}) can be extended to
the limit case $s=0$, so that the Chernoff degree of polarization has
the general expression
\begin{equation}
\mathbb{P}_{\rm C}(\hat \rho)=1-\min_{s\in [0,1]}\left[ \sum_{N=0}^\infty p_N
\left(\xi_N^{(s)}\right)^{1/s}(N+1)^{1-\frac{1}{s}} \right]^s. \label{PC}
\end{equation}

It is well known \cite{Acin,Auden} that the Chernoff overlap is closely
related to the Uhlmann fidelity. This suggests that a comparison between
the Chernoff degree of polarization and the one based on the Bures distance
would be interesting. The Bures degree of polarization has been defined
in Refs. \cite{Bjork2,Bjork4} as
\begin{equation}
\mathbb{P}_{\rm B}(\hat \rho):= 1- \max_{\hat \sigma \in {\cal U}}
\sqrt{{\cal F}({\hat \rho}_b, \,{\hat \sigma}_b)}\,, \label{b0}
\end{equation}
where ${\cal F} $ is the fidelity between two states,
\begin{equation}
{\cal F}({\hat \rho}_1,\,{\hat \rho}_2):=\left[\mbox{Tr}
\sqrt{{\hat \rho}_1^{1/2}{\hat \rho}_2\,{\hat \rho}_1^{1/2}}\right]^2.
\end{equation}
Owing  to the commutation relation (\ref{com}) the following identity holds:
\begin{equation}
[{\cal F}({\hat \rho}_b, \,{\hat \sigma}_b)]^{1/2}
=Q_{1/2}({\hat \rho}_b, \,{\hat \sigma}_b).  \label{f}
\end{equation}
We take advantage of Eq.\ (\ref{f}) to specialize Eq. (\ref{par}) for
the closest unpolarized state,
\begin{equation}
\tilde \pi_N^{(1/2)}=\frac{p_N(N+1)^{-1} \left[\xi_N^{(1/2)}\right]^{2}}
{\sum_{M=0}^\infty p_M (M+1)^{-1}\left[\xi_M^{(1/2)}\right]^{2}},
\label{f1}
\end{equation}
and Eq. (\ref{explicit}) to write the maximal fidelity
${\cal F}({\hat \rho}_b, \,{\hat{\tilde \sigma}}_b)$:
\begin{equation}
{\cal F}({\hat \rho}_b, \,{\hat{\tilde \sigma}}_b)
=\sum_{N=0}^\infty \frac{p_N}{N+1}\left[\xi_N^{(1/2)}\right]^{2}. \label{f2}
\end{equation}
Therefore, the Bures degree of polarization (\ref{b0}) has the expression
\cite{Bjork4}
\begin{equation}
\mathbb{P}_{\rm B}(\hat \rho) = 1- \sqrt{\sum_{N=0}^\infty \frac{p_N}{N+1}
\left[\xi_N^{(1/2)}\right]^{2}}. \label{PB}
\end{equation}

We stress that the polarization measures $\mathbb{P}_{\rm C}(\hat \rho)$,
Eq.\ (\ref{PC}), and $\mathbb{P}_{\rm B}(\hat \rho)$, Eq.\ (\ref{PB}),
depend only on the photon-number probabilities $p_N$ and on the eigenvalues
$\lambda_{N,n}$ of the $N$-photon density matrices 
$\frac{1}{p_N}\, {\rho}^{(N)},\; (p_N>0).$ Hence both of them are nice examples 
for the discussion at the end of Sec. II. Note finally the inequality
\begin{equation}
\mathbb{P}_{\rm C}(\hat \rho) \geqq \mathbb{P}_{\rm B}(\hat \rho).
\label{PC>PB}
\end{equation}

\section{Applications}

\subsection{Pure states}

Let us now analyze the case of a pure state,
$\hat \rho =|\Psi \rangle \langle\Psi|$:
\begin{equation}
|\Psi \rangle =\sum_{N=0}^\infty \sum_{n=0}^Nc_{N,n}\,|n,N-n \rangle,
\qquad \sum_{N=0}^\infty \sum_{n=0}^N|c_{N,n}|^2=1.  \label{pure}
\end{equation}
Its block-diagonal part is a convex combination of $N$-photon pure states,
\begin{equation}
\left[|\Psi \rangle \langle\Psi|\right]_b=\sum_{N=0}^\infty
p_N |\Psi^{(N)} \rangle \langle\Psi^{(N)}|,  \label{rhobpure}
\end{equation}
which is expressed in terms of the photon-number probabilities
\begin{equation}
p_N=\sum_{n=0}^N|c_{N,n}|^2  \label{pNpure}
\end{equation}
and the $N$-photon state vectors
\begin{equation}
|\Psi^{(N)} \rangle :=\frac{1}{\sqrt{p_N}}\sum_{n=0}^Nc_{N,n}\,|n,N-n \rangle,
\qquad p_N>0.  \label{PsiN}
\end{equation}
Each $N$-photon pure state
${\hat \rho}_N=|\Psi^{(N)}\rangle \langle\Psi^{(N)}|$ entering the convex decomposition\ (\ref{rhobpure}) has the eigenvalues
$\lambda_{N,n}=\delta_{n0},$ for $n=0,1,...,N$. Accordingly, Eqs.\ (\ref{PC})
and\ (\ref{PB}) simplify to
\begin{equation}
{\mathbb{P}_{\rm C}}(|\Psi \rangle \langle\Psi|)=1-\min_{s\in [0,1]}
\left[\sum_{N=0}^\infty p_N(N+1)^{1-\frac{1}{s}}\right]^s \label{pureC}
\end{equation}
and, respectively,
\begin{equation}
{\mathbb{P}_{\rm B}}(|\Psi \rangle \langle\Psi|)=1-\left(\sum_{N=0}^\infty
\frac{p_N}{N+1}\right)^{1/2}. \label{pureB}
\end{equation}
As already remarked in Ref. \cite{Bjork4}, for a pure state,\\
$\hat \rho =|\Psi \rangle \langle\Psi|$, the Chernoff and Bures degrees
of polarization are determined solely by its photon-number distribution,
regardless of the nature of the $N$-photon state vectors\ (\ref{PsiN}).

We further specialize the above formulas to the case of a pure state
with $N$ photons, ${\hat \rho}_N=|\Psi^{(N)}\rangle \langle\Psi^{(N)}|,$
whose photon-number probabilities are $p_M=\delta_{MN}$. Hence
Eqs.\ (\ref{pureC}) and (\ref{pureB}) reduce to
\begin{equation}
\pc\left( \proj{\Psi^{(N)}}{\Psi^{(N)}}\right)=\frac{N}{N+1}, \label{pureNC}
\end{equation}
since the minimum over $s$ is reached at $\tilde s=0$, and, respectively,
\begin{equation}
{\mathbb{P}_{\rm B}}(|\Psi^{(N)}\rangle \langle\Psi^{(N)}|)=1
-\left(\frac{1}{N+1}\right)^{1/2}. \label{pureNB}
\end{equation}
Both degrees of polarization are strictly increasing functions of $N$ from
the lowest value  ${\mathbb{P}_{\rm C}}={\mathbb{P}_{\rm B}}=0$,
for the vacuum, to the large-photon-number limit
$$\lim_{N\rightarrow\infty}{\mathbb{P}_{\rm C}}
=\lim_{N\rightarrow\infty}{\mathbb{P}_{\rm B}}=1.$$

\subsection{States with a given photon-number distribution}

Let us consider the set of all two-mode states (pure and mixed)
with a given photon-number distribution $\{ p_N\}_{N=0,1,2,3,...}$.
According to Eqs. (\ref{PC}) and (\ref{PB}), such a state  is maximally
polarized  if and only if its block-diagonal part ${\hat \rho}_b$ is
a convex combination\ (\ref{rhob}) of pure $N$-photon states:
\begin{equation}
{\hat \rho}_b= \sum_{N=0}^\infty p_N|\Psi^{(N)}\rangle \langle\Psi^{(N)}|.
\label{allNpure}
\end{equation}
A significant example is that of the pure state
\begin{equation}
|\Psi\rangle=\sum_{N=0}^\infty \sqrt{p_N}\:|\Psi^{(N)}\rangle \label{super}
\end{equation}
that has the property\ (\ref{rhobpure}). Therefore, the maximal Chernoff
and Bures degrees of polarization are those for a pure state, i.e.,
they are given by Eqs.\ (\ref{pureC}) and\ (\ref{pureB}), respectively.

In what follows we analyze two families of states, each of them having
only two nonvanishing $N$-photon probabilities. The first one is
a one-parameter family of pure states, while the second one consists of
Fock-diagonal mixed states.

\subsubsection{Superposition of two pure $N$-photon states}

Suppose that $N_1$ and $N_2$ are fixed numbers of photons, and $N_1< N_2$.
We investigate the family of pure states
\begin{equation}
\ket{\Psi}=\sqrt{p}\,|\Psi^{(N_1)}\rangle +\sqrt{1-p}\,|\Psi^{(N_2)}\rangle,
\label{vector}
\end{equation}
depending on the probability $p\in [0,1]$.
The block-diagonal part\ (\ref{rhobpure}) of a given state is
\begin{equation}
\left[|\Psi \rangle \langle \Psi |\right]_b
=p\,|\Psi^{(N_1)}\rangle \langle \Psi^{(N_1)}|
+(1-p)\,|\Psi^{(N_2)}\rangle \langle \Psi^{(N_2)}|,
\end{equation}
so that the R\'enyi overlap $Q_s({\hat \rho}_b, \,{\hat \sigma}_b)$,
Eq.\ (\ref{Renyi2}), reads
\begin{equation}
Q_s(p,\pi_{N_1})=p^s\left( \frac{\pi_{N_1}}{N_1+1}\right)^{1-s}
+(1-p)^s \left( \frac{1-\pi_{N_1}}{N_2+1}\right)^{1-s}.
\label{q-tau1}
\end{equation}
In the limit cases $p=0$ and $p=1$, the state vector (\ref{vector})
reduces to $|\Psi^{(N_2)}\rangle$ and $|\Psi^{(N_1)}\rangle$, respectively.
According to Eqs. (\ref{pureNC}) and (\ref{pureNB}), we write
\beqa
&&{\mathbb{P}_{\rm C}}(|\Psi \rangle \langle\Psi|)=\frac{N_2}{N_2+1}, \hspace{0.2cm}
{\mathbb{P}_{\rm B}}(|\Psi \rangle \langle\Psi|)=1-\frac{1}{\sqrt{N_2+1}},\nonumber \\
&&p=0,  \label{pureN2}
\eeqa
and
\beqa
&&{\mathbb{P}_{\rm C}}(|\Psi \rangle \langle\Psi|)=\frac{N_1}{N_1+1}, \hspace{0.2cm}
{\mathbb{P}_{\rm B}}(|\Psi \rangle \langle\Psi|)=1-\frac{1}{\sqrt{N_1+1}},\nonumber \\
&& p=1.   \label{pureN1}
\eeqa
In the case $0<p<1$, it is convenient to write the optimal value (\ref{par})
of the parameter $\pi_{N_1}$,
\begin{equation}
\tilde \pi_{N_1}^{(s)}=\left[1+\frac{1-p}{p}\left(\frac{N_1+1}{N_2+1}\right)
^{\frac{1}{s}-1}\right]^{-1},   \label{pimax}
\end{equation}
as well as the maximum over $\pi_{N_1}$, Eq.\ (\ref{explicit}),
of the R\'enyi overlap (\ref{q-tau1}),
\begin{equation}
Q_s(p, \tilde \pi_{N_1}^{(s)})=\left[ p(N_1+1)^{1-\frac{1}{s}}
+(1-p)(N_2+1)^{1-\frac{1}{s}}\right]^s. \label{maxQ_s}
\end{equation}
 By use of Eqs. (\ref{pureC}) and (\ref{pureB}), we get
\beqa
{\mathbb{P}_{\rm C}}(|\Psi \rangle \langle\Psi|)&=&1-\min_{s\in [0,1]}
\bigg[ p(N_1+1)^{1-\frac{1}{s}}\nonumber\\
&&+(1-p)(N_2+1)^{1-\frac{1}{s}}\bigg]^s,
\label{pC}
\eeqa
and, respectively,
\begin{equation}
{\mathbb{P}_{\rm B}}(|\Psi \rangle \langle\Psi|)=1-\left(\frac{p}{N_1+1}
+\frac{1-p}{N_2+1}\right)^{1/2}. \label{pB}
\end{equation}
The Bures degree of polarization (\ref{pB}) strictly decreases with
the probability $p$.

We are left to find the minimum over $s$ in Eq.\ (\ref{PC}).
The necessary condition for minimum reduces to the transcendental equation
\begin{eqnarray}
&&p(N_1+1)^{1-\frac{1}{\tilde s}}\; \ln \bigg( (N_1+1)\bigg[ p(N_1+1)^{1-
\frac{1}{\tilde s}}\nonumber \\
&&+(1-p)(N_2+1)^{1-\frac{1}{\tilde s}}\bigg] ^{\tilde s}\bigg)+\nonumber \\
&&(1-p)(N_2+1)^{1-\frac{1}{\tilde s}}\; \ln \bigg( (N_2+1)\bigg[ p(N_1+1)^{1-
\frac{1}{\tilde s}}\nonumber \\
&&+(1-p)(N_2+1)^{1-\frac{1}{\tilde s}}\bigg] ^{\tilde s}\bigg)
= 0.  \label{trans}
\end{eqnarray}
Equation\ (\ref{trans}) has no solution for $p\ge \frac{1}{N_1+1}$,
when there is no saddle point of the R\'enyi overlap (\ref{q-tau1}).
The minimum over $s$ in Eq. (\ref{PC}) is reached in $\tilde s=0$.
Further, Eqs. (\ref{pimax}) and (\ref{maxQ_s}) give $\tilde \pi_{N_1}=1$
and $\tilde Q=\frac{1}{N_1+1}$, respectively. The Chernoff degree
of polarization is independent of the probability $p$:
\begin{equation}
{\mathbb{P}_{\rm C}}(|\Psi \rangle \langle\Psi|)=\frac{N_1}{N_1+1},
\qquad\qquad  \frac{1}{N_1+1}\leq p<1 .  \label{plateau}
\end{equation}
In the opposite situation, $p<\frac{1}{N_1+1}$, Eq.\ (\ref{trans})
has a solution $\tilde s\in(0, 1)$. This corresponds to a saddle point
of the R\'enyi overlap (\ref{q-tau1}). The Chernoff degree of polarization
\ (\ref{PC}) depends on the probability $p$, taking values in the interval
\begin{equation}
{\mathbb{P}_{\rm C}}(|\Psi \rangle \langle\Psi|) \in
\left(\frac{N_1}{N_1+1},\; \frac{N_2}{N_2+1}\right),
\hspace{0.2cm}  0<p<\frac{1}{N_1+1}.  \label{variable}
\end{equation}
The above analysis is illustrated in Fig. 1 for a superposition
with lower photon numbers at a fixed value of the probability $p$.
The numerical calculation of the Chernoff degree of polarization
by the saddle-point method is straightforward and can be performed
with great accuracy. Figure 2 displays the comparison between the
maximal (pure-state) Chernoff and Bures degrees of polarization
as functions of the probability $p$.
\begin{figure}
\center
\includegraphics[width=8cm]{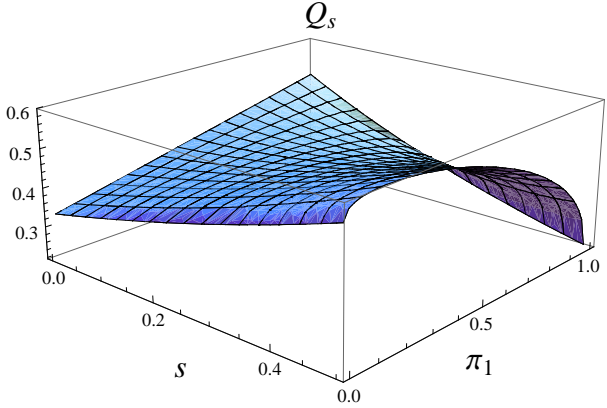}
\caption{(Color online) Displaying the saddle-point evaluation of the Chernoff 
degree of polarization ${\mathbb{P}_{\rm C}}(|\Psi \rangle \langle\Psi|)$
for a state (\ref{vector}) with $N_1=1$, $N_2=2$, and $p=0.1$.
The R\'enyi overlap,  Eq.\ (\ref{q-tau1}), is plotted vs $s$ and $\pi_1$.
The saddle point has the coordinates 
$\tilde s=0.124$ and ${\tilde \pi}_1^{(\tilde s)}=0.634$.
The Chernoff overlap, Eq.\ (\ref{maxQ_s}), is $\tilde Q=0.431$, so the degree
of polarization is ${\mathbb{P}_{\rm C}}=0.569$.}
\label{fig-1}
\end{figure}

\begin{figure}
\center
\includegraphics[width=8cm]{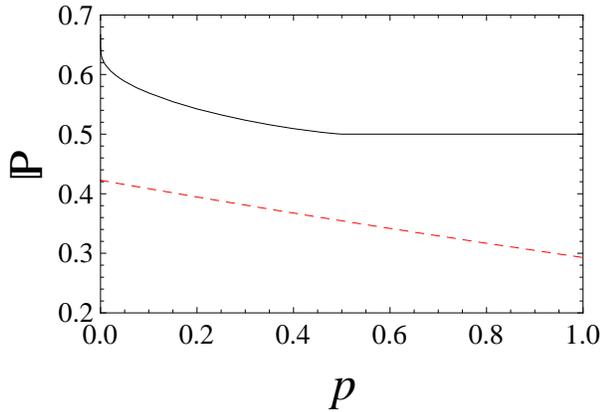}
\caption{(Color online) Degree of polarization of the pure states (\ref{vector})
characterized by $N_1=1$, $N_2=2$ as a function of the probability $p$: 
the Chernoff measure (black full line) and
the Bures measure (red dashed line).}
\label{fig-2}
\end{figure}

\subsubsection{Mixture of two mixed $N$-photon states}

We consider again a pair of fixed numbers of photons, $N_1$ and $N_2$,
such that $N_1< N_2$, and examine a mixture
\begin{equation}
\hat \tau=p\:{\hat \rho}_{N_1}+(1-p)\:{\hat \rho}_{N_2},
\label{mixed}
\end{equation}
where the states ${\hat \rho}_{N_1}$ and ${\hat \rho}_{N_2}$ are Fock-diagonal.
Obviously, ${\hat \tau}_b=\hat \tau.$ In the particular case
when $N_1=1$ and $N_2=2$, we choose density matrices
$\frac{1}{p}\, {\rho}^{(1)}$ and $\frac{1}{1-p}\, {\rho}^{(2)}$ with nonvanishing diagonal entries:
\begin{eqnarray}
\frac{1}{p}\, {\rho}^{(1)}&=&\left( \begin{array}{cc}
\alpha&0\\
0&1-\alpha
\end{array}\right),\nonumber \\
\frac{1}{1-p}\, {\rho}^{(2)}&=&\left( \begin{array}{ccc}
\beta&0&0\\
0&\gamma &0\\
0&0&1-\beta -\gamma
\end{array}\right).
\end{eqnarray}
The R\'enyi overlap (\ref{Renyi2}) specializes to
\beqa
&&Q_s(p, \pi_1)=\left( \frac{\pi_1}{2}\right)^{1-s}p^s\left[ \alpha^s+(1-\alpha)^s\right]\nonumber \\
&&+\left( \frac{1-\pi_1}{3}\right)^{1-s}(1-p)^s\left[ \beta^s+\gamma^s
+(1-\beta-\gamma)^s\right].\nonumber \\
\label{q-tau2}
\eeqa
The Chernoff degree of polarization, Eq. (\ref{PC}), reads
\beqa
\pc(\hat \tau)&=&1-\min_{s\in [0,1]}\bigg\{2^{1-1/s} p\bigg[ \alpha^s
+(1-\alpha)^s\bigg]^{1/s}\nonumber \\
&&+3^{1-1/s}(1-p)\left[ \beta^s+\gamma^s
+(1-\beta-\gamma )^s\right]^{1/s} \bigg\}^s.\nonumber \\
\eeqa
We further write the Bures measure of polarization, Eq. (\ref{PB}):
\beqa
{\mathbb{P}_{\rm B}}(\hat \tau)&=&1-\bigg\{\frac{p}{2}\left[ \alpha^{1/2}
+(1-\alpha)^{1/2}\right]^2\nonumber \\
&&+\frac{1-p}{3}\left[ \beta^{1/2}+\gamma^{1/2}
+(1-\beta-\gamma)^{1/2}\right]^2 \bigg\}^{1/2}. \nonumber \\
\label{Bures2}
\eeqa
Figure 3 presents the saddle-point evaluation of the Chernoff degree
of polarization ${\mathbb{P}_{\rm C}}(\hat \tau)$ of a state (\ref{mixed})
with lower photon numbers. For the same family of states, a comparison
between the Chernoff and Bures degrees of polarization as functions
of the mixing parameter $p$ is made in Fig. 4. Unlike the couple of maximal
degrees of polarization drawn in Fig. 2, their graphs are here very close.

\begin{figure}
\center
\includegraphics[width=8cm]{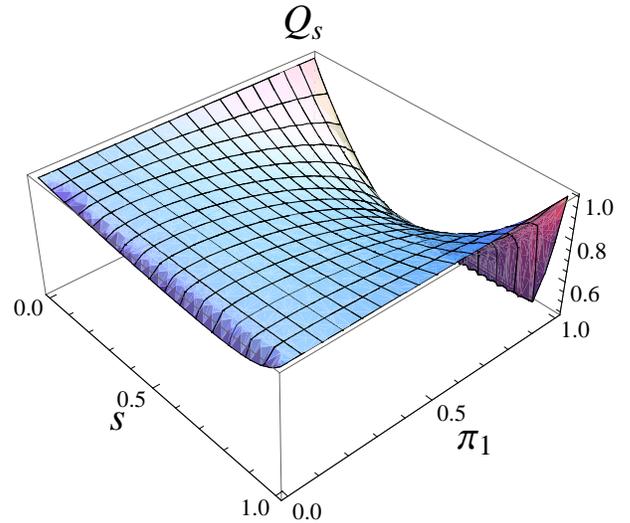}
\caption{(Color online) Saddle-point evaluation of the Chernoff degree
of polarization ${\mathbb{P}_{\rm C}}(\hat \tau)$ for a state (\ref{mixed})
with $N_1=1$, $N_2=2$, $p=0.1$, $\alpha=0.1, \beta=0.01$ and $\gamma=0.04$.
The R\'enyi overlap $Q_s$, Eq.\ (\ref{q-tau2}), is plotted vs $s$ and
$\pi_1$.
The saddle point is reached at  $\tilde s=0.434$ and 
${\tilde \pi}_1^{(\tilde s)}=0.209$. The optimal value $\tilde Q$ is 0.544, 
so that the degree of polarization
is ${\mathbb{P}_{\rm C}}(\hat \tau)=0.251$. For the same state, the Bures 
degree of polarization, Eq.\ (\ref{Bures2}), is 
${\mathbb{P}_{\rm B}}(\hat \tau)=0.247$.}
\label{fig-3}
\end{figure}

\begin{figure}
\center
\includegraphics[width=8cm]{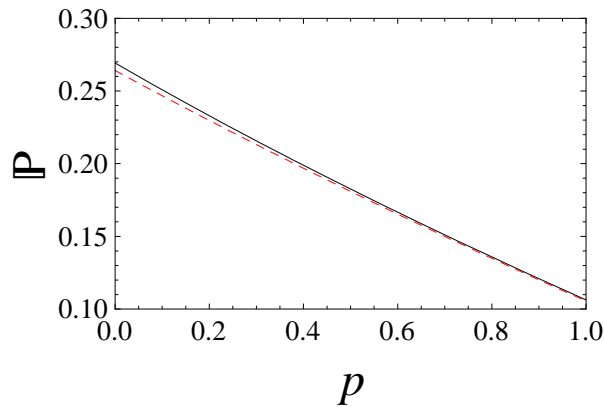}
\caption{(Color online) Degree of polarization of the mixed states (\ref{mixed})
characterized by the same parameters $N_1, N_2, \alpha, \beta, \gamma$ as in Fig. 3 vs the mixing coefficient $p$: the Chernoff measure (black full line) and
the Bures measure (red dashed line).}
\label{fig-4}
\end{figure}

\section{Summary and conclusions}
In this paper we have exploited the quantum Chernoff bound in order to introduce
a distance-type polarization measure for the quantum radiation field.
This measure fulfills the requirements for a genuine degree of polarization,
put forward quite recently \cite{Luis3, Bjork4}. We have derived a general
expression of the Chernoff degree of polarization, Eq.\ (\ref {PC}),
that allows its computation. Moreover, a comparison between the Chernoff
and Bures degrees of polarization proved to be very useful. For instance,
Fig. 2 displays both degrees of polarization for a one-parameter family
of pure states that are superpositions of a fixed pair of pure $N$-photon
states. The Bures polarization measure distinguishes between all the states
of this family because it is strictly decreasing with the probability
of one of the $N$-photon states. On the contrary, the predicted existence
of a plateau of the Chernoff degree of polarization starting from a threshold
of the same probability is displayed. Although considerably larger than
the Bures polarization measure, the Chernoff measure cannot discriminate
between the corresponding states. On the other hand, Fig. 4 points out that
for a one-parameter mixture of two given mixed $N$-photon states, the Bures
and Chernoff degrees of polarization happen to be very close.

We stress that the R\'enyi overlaps $Q_s({\hat \rho}_b, 
{\hat{\tilde \sigma}}_b)$, \hspace{0.85cm} with $0<s<1$, Eq.\ (\ref{explicit}), can themselves
be employed as reliable measures of polarization. The symmetric one
$(s=\frac{1}{2})$ yields the Bures degree of polarization via Eq.\ (\ref{f})
and has a privileged position owing to its significant meaning in quantum
mechanics. To conclude, the Chernoff polarization measure, Eq.\ (\ref{PC}),
deserves special attention because it is the maximal R\'enyi distance-type
polarization measure.

\section*{Acknowledgments}
This work was supported by the Romanian Ministry of Education and Research
through Grant IDEI-995/2007 for the University of Bucharest, the Swedish Foundation
for International Cooperation in Research and Higher Education (STINT),
and the Swedish Research Council (VR).

\end{document}